# A molecular mechanism for the water-hydroxyl balance during wetting of TiO$_2$

M. Amft[1,2], L. E. Walle[3], D. Ragazzon[4], A. Borg[3], P. Uvdal[5], N. V. Skorodumova[1,4], and A. Sandell[4,*]

[1] *Applied Materials Physics, Department of Materials Science and Engineering, Royal Institute of Technology, SE-100 44 Stockholm, Sweden*
[2] *Forsmarks Kraftgrupp AB, SE-74203 Östhammar, Sweden*
[3] *Dept. of Physics, Norwegian University of Science and Technology (NTNU), NO-7491 Trondheim, Norway*
[4] *Dept. of Physics and Astronomy, Uppsala University, P. O. Box 516, SE-751 20 Uppsala, Sweden*
[5] *Chemical Physics, Dept. of Chemistry, P.O. Box 124, and MAX-IV laboratory, P.O. Box 118, Lund University, SE-221 00 Lund, Sweden*

**Abstract**

We show that the formation of the wetting layer and the experimentally observed continuous shift of the H$_2$O-OH balance towards molecular water at increasing coverage on a TiO$_2$(110) surface can be rationalized on a molecular level. The mechanism is based on the initial formation of stable hydroxyl pairs, a repulsive interaction between these pairs and an attractive interaction with respect to water molecules. The experimental data are obtained by synchrotron radiation photoelectron spectroscopy and interpreted with the aid of density functional theory calculations and Monte Carlo simulations.

**Keywords:**

Titanium dioxide; Water; Adsorption; Photoelectron spectroscopy; Density Functional Theory calculations

# 1. INTRODUCTION

The formation of the interface between water and a solid is both a technologically important and challenging fundamental problem. On many metals, metal oxides, semiconductors and minerals wetting of the surface involves water dissociation into hydroxyl groups.[1-5] The hydroxyls play an important role in stabilizing the wetting layer[2,6-8] and are essential for the mechanisms of metal corrosion,[9] molecular diffusion[10] and proton mediated ion transport.[11] A particularly versatile and important material is titanium dioxide ($TiO_2$),[12] for which the nature of the wetting layer is essential for applications such as photovoltaics,[13] photocatalysis[14,15] and photogeneration of hydrophilic coatings.[16] The $TiO_2(110)$ surface of the rutile polymorph is a prototypical substrate not only for studies of $TiO_2$ surface chemistry but also for development of methods to explore metal oxide surface chemistry in general.[17] The stoichiometric $TiO_2(110)$ surface exhibits troughs of fivefold-coordinated Ti atoms [Ti(5)] and rows of twofold-coordinated, bridge-bonded O atoms [O(2)], Figure 1A. Water adsorbs most strongly on titanium atoms and the first water layer (1 monolayer, ML) is defined as completed when all the Ti(5) sites are occupied. For many years, the experimental results were interpreted such that dissociation only occurs at bridging oxygen vacancy defects while no dissociation occurs on the stoichiometric surface.[17] In contrast, theoretical studies based on density function theory (DFT) predicted facile dissociation on the stoichiometric surface at coverages below 0.25 ML.[18-20] A few years ago a recipe for the preparation of defect free $TiO_2(110)$ terraces was presented.[21] We exploited this method in a core level photoemission study to demonstrate that the first layer of water on stoichiometric $TiO_2(110)$ is partially dissociated.[3] Consistently, water dissociation on the stoichiometric surface was proposed in a recent photoelectron diffraction study.[22]

The appropriate theoretical modeling of the system has in the meantime remained the subject of an intense debate.[19,23] The most recent DFT calculations, however, all shows dissociative adsorption at low coverages (≤0.25 ML) while a molecular or a partially dissociated water layer are favored at high coverages (≥0.5 ML).[20,24,25]. These findings are independent of the approximation to the exchange-correlation functional, PW91, RPBE and B3LYP. That is, the experimental and theoretical results are converging but a comprehensive description of the formation of the first water layer is still lacking.

Here we report on the formation of the wetting layer on the $TiO_2(110)$ surface, from zero to monolayer coverage, unifying experiment and theory. We have monitored the growth of the first water layer in ultrahigh vacuum (UHV) by core level photoelectron spectroscopy (PES), obtaining a complete picture of the H₂O-OH partitioning as a function of coverage. Exploiting density functional theory (DFT) calculations and Monte Carlo (MC) simulations, we derive a molecular mechanism for the initial wetting of the $TiO_2$(110) surface. Importantly, the experimental data reported here provide solid information on the chemical identity (H₂O vs OH) of the adsorbed species. Since also the coverages studied experimentally are in perfect correspondence to those in our DFT treatment the integrity of the theoretical efforts can be tested to a very high degree.

## 2. EXPERIMENTAL METHODS

The experiments were performed at beamline D1011 at the Swedish synchrotron radiation source MAX II.[26] The formation of the first layer during water uptake on stoichiometric $TiO_2$(110) was monitored by a set of photoelectron spectra of the O 1s level, recorded at 610 eV photon energy and 60° off normal emission to enhance the surface sensitivity. The stoichiometric $TiO_2$(110) surface was prepared at room temperature (RT) by exposing a sputtered and annealed surface to H₂O, by which OH pairs form at oxygen vacancy sites. In a subsequent step the capping H atoms were titrated away by $O_2$.[3,21,27] On this surface water does not adsorb in any form at RT, giving clear evidence for the absence of oxygen vacancies.[3,21] The present results refer to D₂O adsorption, as D₂O is expected to be even less susceptible than H₂O towards radiation-induced dissociation.[7] However, radiation damage is not an issue in the present study since: (1) the same results were in fact obtained for H₂O, i.e. there is no isotope effect;[7] (2) no time effects were observed and (3) the same monolayer composition was obtained when using three different preparation procedures (slow growth at 210 K, extensive dosing at 210 K and heating of a multilayer to 210 K). The absence of radiation effects is consistent with the low photon doses employed (<0.3 ph/molecule).[28]

## 3. THEORETICAL METHODS

The scalar-relativistic *ab-initio* density functional theory calculations were performed using the projector-augmented wave (PAW) as implemented in VASP.[29] The exchange-correlation interaction was treated in the generalized gradient approximation (GGA) in the parameterization of Hammer, Hansen, and Nørskov (RPBE).[30] The 3d, 4s Ti states and 2s, 2p

O states were treated as valence states. A cut-off energy of 700 eV was used and a Gaussian smearing with a width of σ = 0.05 eV for the occupation of the electronic levels. Spin-polarization was taken into account in all calculations.

The rutile $TiO_2$(110) surfaces were modeled by 2x5 symmetric slabs containing seven O-Ti-O trilayers (420 atoms). The molecules were adsorbed symmetrically on both sides of the slab to avoid any dipole moment across the slab. The lowest coverage (8.3%) was modeled by adsorption of single $H_2O$ molecules on a 2x6 symmetric slab also containing seven O-Ti-O trilayers (504 atoms). The lattice parameters, a = 4.692 Å and c = 2.97 Å, were obtained from the respective bulk calculations. The repeated $TiO_2$ slabs were separated from each other by at least 17 Å of vacuum. Γ-centered Monkhorst-Pack k-point meshes of 3x2x1 (3 k-points) were used for the unit cells. During the structural relaxation the Ti and O atoms in the central layer were kept fixed, while the rest of the structures were free to relax. The relaxation cycle was stopped when the Hellmann-Feynman forces had become smaller than $1\cdot10^{-2}$ eV/Å. The adsorption energies of molecular (dissociated) water molecules ($H_2O^{mol\ (diss)}$) on the $TiO_2$ surface were calculated as $E_{ads} = E(H_2O^{mol\ (diss)}/TiO_2) - E(H_2O^{gas\ phase}) - E(TiO_2)$ and are negative when the adsorption is exothermic. In the Monte Carlo simulations a 1000x1000 mesh with periodic boundary conditions was used to simulate a surface comprising $10^6$ Ti(5f) sites.

## 4. RESULTS AND DISCUSSION

The experimental results are summarized in Figs. 1B-D. The O 1s spectra were fitted as components from $TiO_2$, OD and $D_2O$ as demonstrated in Figure 1B. The recording time for each spectrum was 3 minutes. The sample temperature was kept at 210 K to prevent the formation of a second water layer.[31] A complete map of the O 1s intensity distributions between $TiO_2$, OD and $D_2O$ as a function of time is shown in Figure 1C. Figure 1D shows the resulting $D_2O$ and OD coverages as a function of time. Included is also the total $D_2O$ uptake, obtained by adding the O 1s intensities as [O 1s($D_2O$)] + 0.5[O 1s(OD)] since the dissociation of one water molecule upon adsorption results in two OD groups. The results show that dissociative adsorption is favored in the low coverage limit (<0.2 ML) whereas molecular adsorption dominates above 0.4 ML. Saturation is reached after approximately 10 scans or 30 minutes. At this point, the $D_2O$ coverage is 0.80 (±0.05) ML and the OD coverage is 0.40 (±0.05) ML, i.e. 1 ML uptake. In contrast, at the lowest water uptake measured, 0.06 (±0.01)

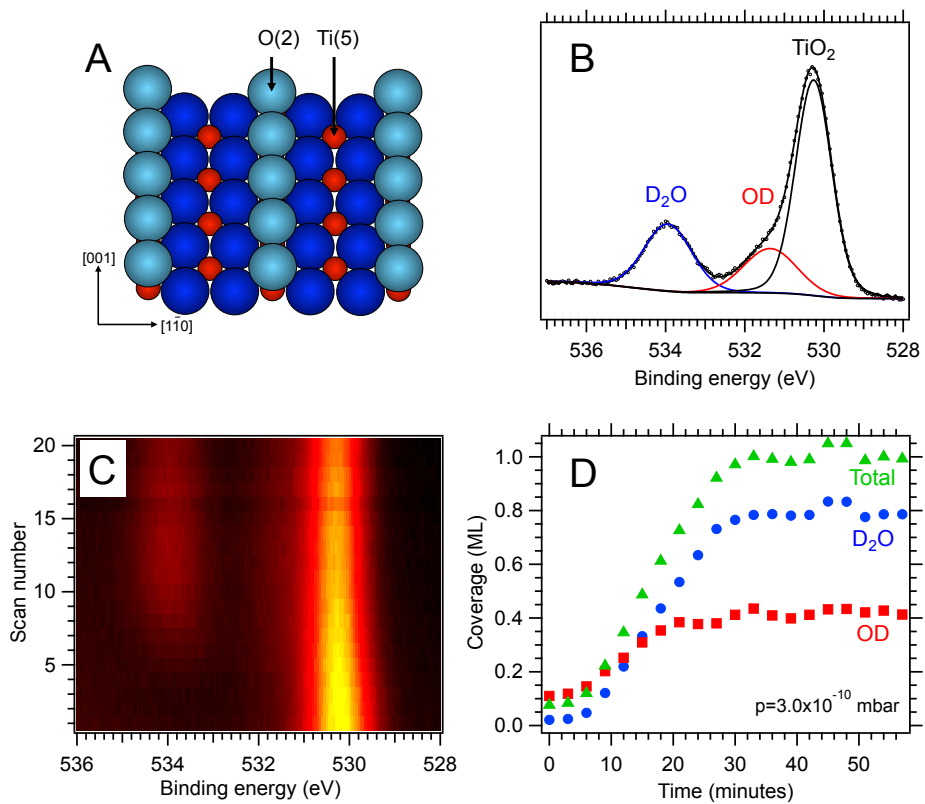

*FIG. 1 (A) Schematic illustration of the rutile $TiO_2$ surface, showing the rows of five-fold coordinated Ti ions [Ti(5)] and bridging O ions [O(2)] defining the sites active for adsorption. (B) An O 1s spectrum for $D_2O/TiO_2(110)$, showing the delineation into components stemming from $TiO_2$, OD and $D_2O$, respectively. (C) A 2D map of the O 1s spectral region vs. time during uptake of $D_2O$. (D) The resulting OD and $D_2O$ coverages (in monolayers, ML).*

ML, $D_2O$ has partitioned into 0.03 (±0.01) ML $D_2O$ and 0.08 (±0.02) ML OD (attributed to uptake from the background during the cooling of the sample). Our data thus show that the $OD/D_2O$ ratio continuously decreases all the way to the full monolayer and that both adsorption states are formed up to about 0.8 ML.

To unveil the molecular mechanisms of the wetting layer growth we turn to the results of the DFT calculations. The 2x6 supercell is considered to represent isolated molecules and 8.3 % is in fact the lowest coverage calculated so far.[20] The 2x5 supercell was employed to study the interaction between molecules on the surface. To simulate the increasing coverage during growth, we filled both surfaces of the 2x5 slab with water in steps of 0.1 ML, in fact, adding molecules one by one and neglecting adsorbate diffusion. When a molecule was added various molecular configurations for this coverage were investigated in order to find the one with the lowest energy. This configuration was used subsequently when adding the next molecule to the surface. Thus, we successively filled the surface with molecules up to full monolayer coverage collecting on the way information about possible structures for each of the considered coverages. This approach allowed us to interpret the basic mechanisms behind the water layer formation and the structure of the resulting monolayer.

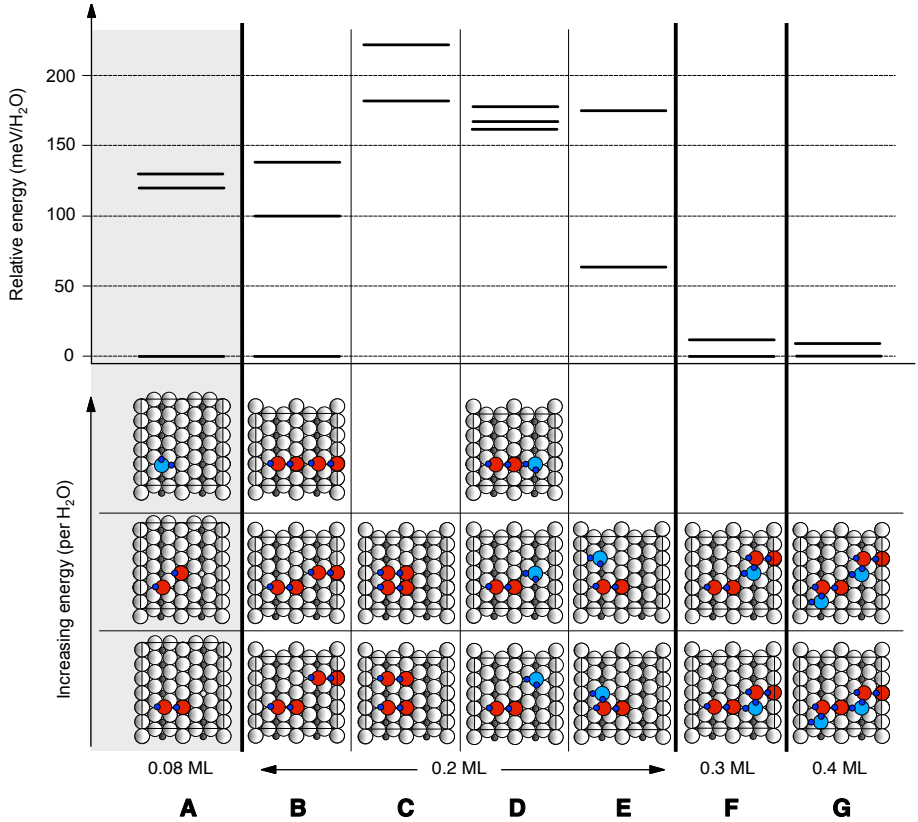

FIG. 2. Bottom panel: Molecular configurations for different water coverages and their relative energies (in meV/$H_2O$ molecule). A 2x6 unit cell was used for the lowest coverage of 0.08 ML (**A**) while a 2x5 unit cell was used for the higher coverages, 0.2 – 0.4 ML (**B-G**). Top panel: Diagram showing the relative energies for the different coverages (in meV/$H_2O$ molecule). The zero energy level represents the lowest-energy configuration for each of the coverages.

The most important results of this investigation are summarized in Figures 2 and 3. In the low coverage limit (0.08 ML), we found a clear preference for water dissociation, i.e. a OH pair is formed, Figures 2A and 3. This pair is stable as the OHs, adsorbed at Ti(5) and adjacent O(2) sites, demonstrate an attractive interaction, Figure 2A.

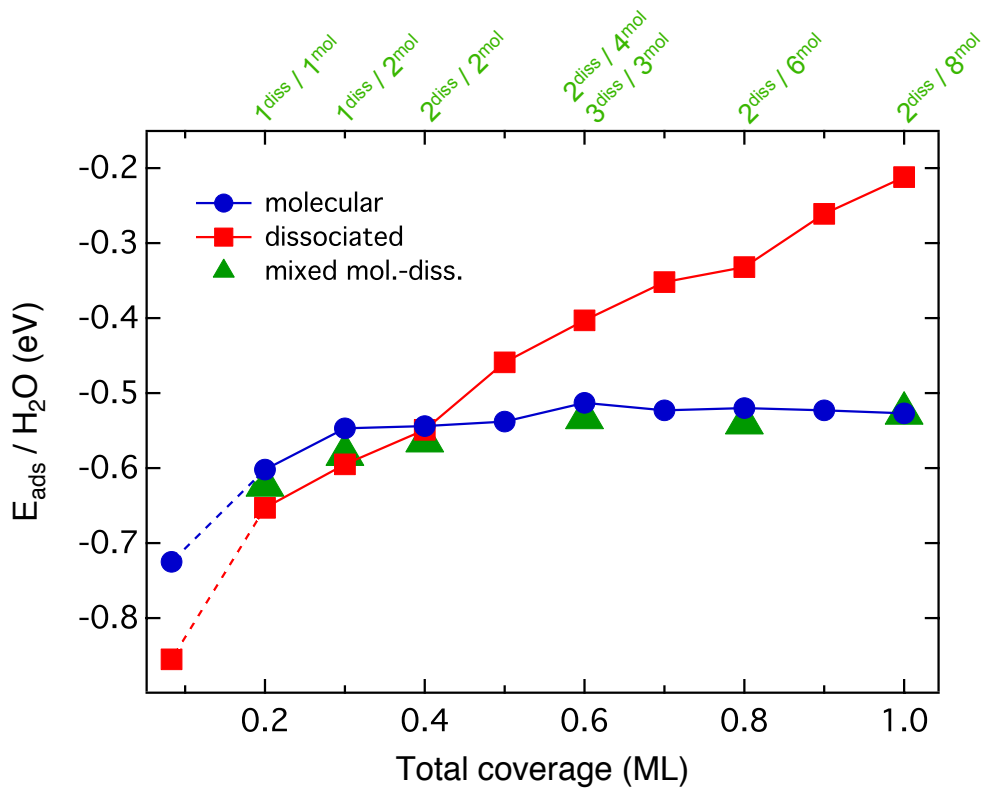

*FIG. 3. Adsorption energies of molecular (circles), dissociated (squares) and mixed molecular-dissociated (triangles) configurations for different coverages. All the reported results were obtained using a 2x5 unit cell except the lowest coverage of 0.08 ML, for which a 2x6 unit cell was used.*

This result rationalizes the absence of isotope scrambling found upon desorption.[32] The interactions between two OH pairs and one OH pair and one $H_2O$ molecule are shown in Figures 2B-E, at to 0.2 ML coverage. The interaction between these OH pairs is repulsive both along the trough and between OH pairs located in neighboring troughs, Figures 2B-C. An attractive interaction is found between an OH pair and a water molecule when they are adsorbed in the same trough, Figure 2D. When they are located in neighboring troughs the differences in interaction energies are minor; there is possibly a small repulsive interaction. The results of the 0.2 ML-configurations show that: (1) A molecule arriving in the neighboring trough to a dissociated molecule minimizes the energy by dissociation on the Ti(5) site displaced (at least) two steps from the OH pair, Figure 2B (bottom). (2) A molecule adsorbing in the same trough as a OH pair needs to be separated by a minimum of two Ti(5) sites in order for dissociation to be energetically favorable. This can be inferred from comparison of the configurations shown in Figures 2C and 2E. The energies become nearly degenerate at this separation and at further separation the configuration with two dissociated molecules will become lowest in energy.

At higher coverages (0.3 and 0.4 ML), Figure 2F and 2G, we observe a clear tendency towards clustering of purely molecular and mixed (molecular and dissociated) configurations. The repulsive interaction between OH pairs is furthermore clearly reduced when water molecules are attached to one or two of the OH pairs, Figure 2F and 2G. That is, higher coverages allow for denser packing of dissociated species.

The origin of the repulsive interaction between the OH pairs is a large distortion of the $TiO_2$ surface upon hydroxylation. Our calculations show that the distortion extends two tri-layers into the substrate. The major distortion is caused by the adsorption of an OH group on a Ti(5) site, inducing an out-of-plane relaxation of the Ti(5) ion by 0.73 Å. Substantial, predominantly vertical, distortions are also found in the surrounding surface area. The adsorption of a molecular $H_2O$ on the next neighboring Ti(5) site to the dissociated species, i.e. the formation of a $H_2O$-(OH-OH) configuration (Figure 2E, bottom), reduces the vertical distortion of the Ti(5) ion under the OH group to 0.50 Å. This reduction in the geometrical distortion is the reason why denser packing of OH pairs at higher coverage becomes possible. Adding more $H_2O$ molecules to the mixed dimer does not decrease the distortion further. In contrast, the adsorption of molecular water itself causes only a small relaxation of the surface (less than 0.01 Å).

Figure 3 summarizes the energies for purely molecular and dissociative adsorption as well as the energies for the optimal mixed configurations, identified in the surface filling procedure described above. The curve for molecular water crosses that for dissociated $H_2O$ at about 0.4 ML coverage. The mixed configurations are energetically more favorable from about 0.3 ML coverage all the way to the monolayer coverage where the energy of the mixed configuration is only 0.5 meV/$H_2O$ lower than that of molecular water, that is they become practically degenerate.[20,24] Thus, our procedure has allowed us to find a number of mixed configurations that give lower energy than those of molecular water adsorption only.

Next, we constructed a straightforward Monte Carlo model to simulate the formation of the wetting layer on the $TiO_2$ surface. The essential input into this simulation is the size of the area prohibited for additional dissociation due to the repulsive interaction between the OH pairs. In the MC runs $H_2O$ molecules were randomly dropped onto the surface, containing $10^6$ Ti(5) sites, i.e. a 1000x1000 mesh with periodic boundary conditions, and their dissociation was allowed outside the repulsive interaction area but forbidden within. Several sizes of this critical area were considered, Figure 4. For lower coverages our DFT results indicate that the dissociative adsorption is not possible on the Ti(5) sites closer than 8.91 Å from an existing OH pair (an effective repulsive interaction area of 217 Å$^2$), blue frame, inset in Figure 4. The OH/H$_2$O ratios obtained by the MC simulation for this case are shown in Figure 4 (blue curve). An excellent agreement with the experimental data is found below 0.4 ML water uptake. The MC curve deviates from the experimental results at higher coverages as it ignores the reduced repulsive interaction between the OH pairs upon water attachment, suggested by our DFT calculations. Shortening the minimum distance between dissociated species in the same trough yields an MC curve that shows a good agreement with the experimental data at higher coverages as well (area 178 Å$^2$, magenta frame and curve in Figure 4). If the size of the effective repulsive interaction area is substantially smaller (138 Å$^2$, black frame and curve) or larger (296 Å$^2$, yellow frame and curve) the MC results clearly deviate from the experimental results.

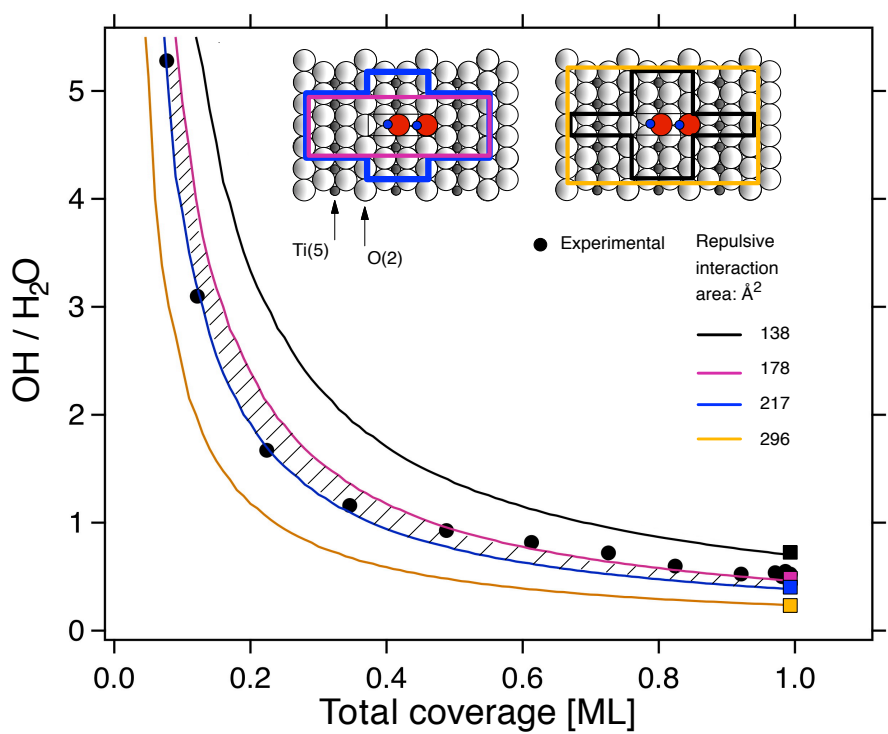

*FIG. 4. Coverage dependences of the OH/H$_2$O ratio obtained in Monte Carlo simulations for different sizes of the repulsive interaction region between OH-OH pairs (lines) as compared to the experimentally measured values (filled circles). The insets illustrate the areas around a dissociated H$_2$O that is prohibited for the adsorption of additional dissociated H$_2$O.*

## 5. CONCLUSIONS

To summarize, our combined experimental and theoretical results reveal that the adsorption of water on $TiO_2(110)$ is governed by a delicate balance between molecular and dissociative adsorption which shifts towards molecular adsorption upon increasing coverage. The balance can be understood by a straightforward mechanism based on the initial formation of stable hydroxyl pairs, a repulsive interaction between these pairs and an attractive interaction with respect to water molecules. The repulsive interaction between the OH pairs is induced by a structural distortion of the substrate lattice. The straightforwardness of the molecular mechanism behind the wetting of the $TiO_2$ surface disclosed in the present work makes it tempting to suggest that it can be generalized to the wetting of other metal oxide surfaces as well where OH is formed.

## ACKNOWLEDGEMENTS

We thank the staff at MAX-lab for their support. Financial support was received from the Swedish Energy Agency (Energimyndigheten, STEM), the Swedish Research Council (Vetenskapsrådet), Research and Innovation for Sustainable Growth (VINNOVA), the Trygger foundation, the Crafoord foundation, NordForsk and the Göran Gustafsson foundation. L.E.W. has been supported through the Strategic Area Materials at NTNU. Supercomputer time was granted by the Swedish National Infrastructure for Computing (SNIC).